\documentclass{article}

\usepackage[a4paper, margin=1.2in]{geometry}
\usepackage{authblk}
\usepackage{cite}

\usepackage{graphicx}
\usepackage{subcaption}
\usepackage{booktabs}
\usepackage{array}
\usepackage{url}
\usepackage{cite}
\usepackage{amssymb}
\usepackage[hidelinks]{hyperref}

\graphicspath{{figures/}}
\newcommand{\code}[1]{\texttt{\small #1}}

\providecommand{\keywords}[1]{\textbf{\textit{Keywords:}} #1}

\title{Configuration, Not Conscience: A Large-Scale Empirical Study of LLM System Prompts}

\author[1,2]{Constantinos Patsakis\thanks{\url{kpatsak@unipi.gr}}}
\author[1]{Vasilios Argyropoulos\thanks{\url{bargiropoulos@unipi.gr}}}
\author[1]{Efthymios Alepis\thanks{\url{talepis@unipi.gr}}}

\affil[1]{Department of Informatics, University of Piraeus, Piraeus, Greece}
\affil[2]{Athena Research Centre, Greece}

\date{}
\begin{document}
\maketitle

\begin{abstract}
Leaked system prompts are often treated as windows into the hidden values of commercial language models, yet their composition is rarely studied at scale. We analyze a merged corpus of 407 leaked, reconstructed, or officially published system prompts from 62 vendors across four community collections, identifying 29 near-duplicate clusters covering 66 files. Operational content rather than ethical statements dominates the corpus; a deliberately simple block-level classifier assigns roughly 58\% of classified words to tool/protocol and roughly 5\% to safety policy, while the strictest rule-lines guard tool use and file safety over harmful content by an 11:1 margin. Literal text transfer concentrates in a small set of cross-vendor pairs. Prompts also carry measurable maintenance debt, with version chains turning over thousands of words per release. The evidence supports treating leaked prompts as operational specifications, closer to configuration files than value statements, and treats reuse and prompt rot as engineering and supply-chain concerns. Because most documents are adversarial in origin and the detectors are deliberately simple, all magnitudes are directional; we audit the main classifier's error modes.
\end{abstract}


\section{Introduction}
The system prompt is the standing instruction layer a deployed language model receives before any
user turn. It fixes the assistant's identity, tools, output format, refusal and escalation behavior,
and safety, legal, and product-specific norms. In an LLM-integrated application, this text is the
main control mechanism available to product teams, yet it is rarely published and
studied directly. Public attention tends to treat leaks as confessions about vendor values,
guardrails, or refusal triggers. Security work has likewise treated system prompts mainly as targets
for extraction, indirect injection, or jailbreak probing, and practitioner taxonomies now treat the
prompt layer as a first-class application risk (Section~\ref{sec:related}). Far less is studied about
what these documents contain, e.g., how their words are distributed across operational, safety, legal, and
stylistic concerns, how much text vendors reuse, and how prompts evolve across releases.

We study these questions with a merged corpus from four independently maintained community
collections of leaked and reconstructed prompts. To this end, we ask: (1) what types of prompts the corpus
contains; (2) how much literal and semantic reuse exists across prompts, vendors, and source
repositories; (3) which operational, safety, legal, formatting, and injection-defense concerns are
most frequent; and (4) how prompt versions evolve in length, content, autonomy, and maintenance
risk.

Our study measures 407 prompts from 62 vendors after merging four repositories and flagging
near-duplicates rather than silently double-counting them. It shows that observed leaked prompt text
is dominated by operational specification rather than value or refusal language, corroborated by a
five-way matched comparison against vendor-published prompts. It separates literal, semantic, and
structural reuse, localizing verbatim copying to a small cluster of coding agents while showing that
the strongest architectural convergence occurs without copied text. It also quantifies prompt
growth, revision volume, and ``rot risk,'' treating prompt maintenance as a software engineering
problem grounded in a vendor-published release-note archive with verified dates.

\section{Related Work}
\label{sec:related}
\textbf{System prompts as a security target.} Prior work mostly treats system prompts as
adversarial targets. Prompt extraction recovers them from deployed
systems~\cite{DBLP:journals/corr/abs-2211-09527,zhang2024effective}; indirect prompt injection subverts them through
untrusted data reaching the model's context~\cite{DBLP:conf/ccs/AbdelnabiGMEHF23,liu2024formalizing}, with agentic
benchmarks~\cite{DBLP:conf/nips/DebenedettiZBB024} and design-level
defenses~\cite{DBLP:journals/corr/abs-2506-08837} following; and jailbreak work probes the refusal behavior
they encode~\cite{DBLP:conf/nips/0001HS23,schulhoff2023ignore}. Practitioner taxonomies have since
promoted the prompt layer to a first-class application risk~\cite{owasp2025top10,owasp2025llm01,
nist2024genai,mitre2026atlas}. These studies ask whether a system prompt can be recovered or
overridden; they do not ask what the recovered documents contain. This paper studies the recovered
texts themselves rather than treating them only as attack outputs.

\textbf{Prompts as engineered documents.} Prompt-engineering research is largely prescriptive,
cataloguing patterns authors should apply~\cite{white2023prompt}, while empirical work shows that
formatting choices affect model behavior~\cite{sclar2024quantifying}; this motivates direct
measurement of captured system-prompt texts. Transparency work asks what vendors should publish,
through model cards~\cite{mitchell2019model}, datasheets~\cite{gebru2021datasheets}, and value
specifications such as constitutional training~\cite{bai2022constitutional}. Our matched
official/leaked comparison (Section~\ref{sec:official}) connects these threads: the published text
reads as a value statement, but is much shorter than the matched leaked text observed here.

\textbf{Methods borrowed from software engineering.} Our reuse analysis adapts shingling with
Jaccard resemblance and asymmetric containment~\cite{broder1997resemblance}, and follows software
clone detection~\cite{kamiya2002ccfinder,DBLP:conf/icmss/MinP19} in separating verbatim copying from
architectural convergence (Section~\ref{sec:structural}). Semantic comparison uses sentence
embeddings~\cite{reimers2019sentence,wang2020minilm}. The prompt-rot analysis
(Section~\ref{sec:version}) treats stale references and contradictory directives as technical
debt~\cite{kruchten2012technical}.

\section{Corpus and Methods}
\label{sec:corpus}
The corpus merges four community-maintained prompt collections, namely: CL4R1T4S~\cite{cl4r1t4s} (commit \code{09916a9}, 66 files),
System-Prompts~\cite{sysprompts} (commit \code{0d04045}, 30 coded files),
system-prompts-and-models-of-ai-tools~\cite{spamt} (commit \code{570364c}, 112 coded files), and
system\_prompts\_leaks~\cite{spleaks} (commit \code{5d3c769}, 231 coded files). After excluding
non-prompt material (tool-call JSON schemas with no natural-language instruction, README/license/index
files, build scripts, and Claude Code's bundled tool/skill documentation, which describes the agent's
own tools rather than constituting its system prompt), the merged corpus contains \emph{407 coded
files from 62 vendors}, collected 2026-06-17 through 2026-07-11.

Because the four repositories are collected independently, the same underlying document is
sometimes harvested by more than one. We detect this with word 5-gram Jaccard within each vendor,
clustering pairs at Jaccard $\geq 0.85$ via connected components, gated by a guard that blocks the
edge when two filenames carry disjoint model-tier/version tokens (an Anthropic Opus/Sonnet same-day
pair can reach Jaccard 0.97 while being two different models, not a re-scrape). This finds 29
duplicate clusters covering 66 files (mean size 2.3, max 4), leaving 370 effectively unique documents;
we keep every file (tagged with a cluster id) rather than discard redundant copies, since
cross-repository agreement is itself evidence of capture fidelity, and report both raw (407) and
deduplicated (370) counts.

Manual labels assign each file to one of six archetypes (chat assistant, coding agent, autonomous
agent, app builder, narrow/embedded assistant, search/research assistant) and one value posture
(operational, caution, truth-seeking), using close reading against a fixed taxonomy (a single-coder
threat to validity discussed in Section~\ref{sec:threats}).
Of the 407 coded files, 236 are operational, 156 caution, and 15 truth-seeking. The latter two labels
concentrate in chat-facing prompts: 152 of 155 chat files carry caution or truth-seeking, as do 19 of
99 narrow fragments that are persona/safety snippets layered on chat prompts (e.g., voice-mode
variants or standalone image-safety-policy blocks). Every coding-agent, autonomous-agent,
app-builder, and search/research prompt is operational. Thus, the documents often treated as ethical
statements are concentrated in conversational prompts; the rest of the corpus is operational by both
manual judgment and lexical measurement.

Most of the corpus is adversarial in origin, as files were obtained through extraction or community
reconstruction, not vendor release, and we cannot verify their fidelity, completeness, or dating. One
subset is different and is tagged accordingly (\code{provenance=official} on 31 of 407
files): system\_prompts\_leaks archives Anthropic's own published release-notes system prompts from
\url{platform.claude.com/docs}, spanning 2024-07-12 through 2026-05-28. These are
vendor-published, not leaked, and we use them in Section~\ref{sec:official} both as a fidelity check
on the leaked majority of the corpus and as a verified-date reference point for the version-drift
analysis in
Section~\ref{sec:version}. For the remaining 376 leaked/reconstructed files, every result describes
the observed documents, not any vendor's true policy or any deployed model's behavior. The analysis
uses only already-public text; we perform no new extraction attacks and publish no private user data.
We report vendor-level aggregates rather than amplifying individual instructions that would help
extraction or evasion. Similarity scores likewise describe document relationships, not whether any
organization copied from or collaborated with another.

\begin{table}[t]
\centering
\caption{Coded corpus composition. Word counts are per file; SD is the standard
deviation across files in the archetype.}
\label{tab:arch}
\scriptsize
\begin{tabular}{lrrr}
\toprule
Archetype & Files & Mean words & SD words \\
\midrule
Chat assistant & 155 & 5{,}630 & 7{,}114 \\
Coding agent & 108 & 3{,}878 & 4{,}975 \\
Autonomous agent & 22 & 4{,}590 & 6{,}906 \\
App builder & 18 & 7{,}153 & 12{,}219 \\
Narrow/embedded & 99 & 1{,}368 & 2{,}218 \\
Search/research & 5 & 989 & 345 \\
\bottomrule
\end{tabular}
\end{table}

\begin{figure}[t]
\centering
\includegraphics[width=0.8\columnwidth]{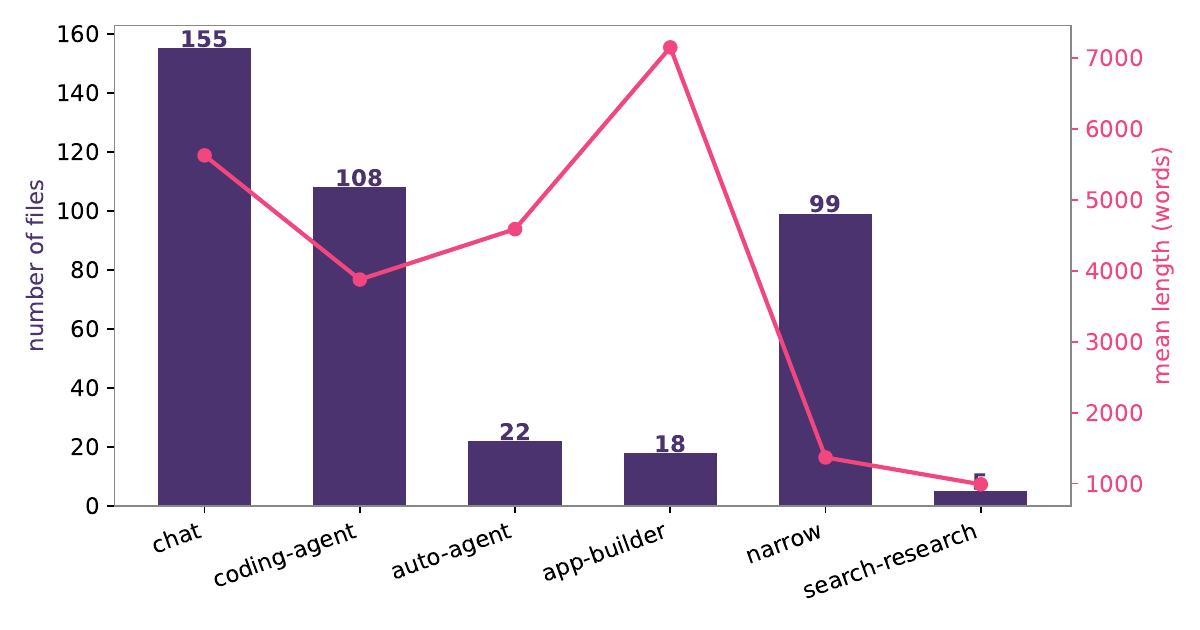}
\caption{Corpus composition by archetype. Bars show file counts and the line shows mean word count.}
\label{fig:arch}
\end{figure}

Regex detectors capture emphasis markers, feature incidence, capability families, harm categories,
injection-defense techniques, formatting requirements, tone markers, legal-risk categories,
hardcoded rot-risk markers, refusal contexts, and contradiction/tension flags. They are simple,
deterministic, and auditable; they may over-count permissive mentions and under-count paraphrases, so
we treat headline percentages as directional. The regexes are English-pattern based. A corpus-wide
scan found no substantively non-English files, only small embedded multilingual examples, so no file
was excluded on language grounds.

The block taxonomy splits prompts into paragraph-like chunks and assigns each to one functional
bucket (strongest lexical match): tool/protocol, safety policy, legal/IP, formatting/style,
identity/persona, memory/context, injection boundary, or other. A mixed-purpose paragraph therefore
contributes wholly to its dominant bucket, understating secondary concerns. Block taxonomy and
vendor/archetype rates use the full 407-file corpus. Version diffing
(Section~\ref{sec:version}) uses word-level sequence matching between consecutive prompts in five
chains: two Anthropic official chains (Sonnet-tier, Opus-tier), one leaked-only Claude chain
(CL4R1T4S), and one each for ChatGPT and Grok. Replacements count as both removal and addition, so
diff counts measure turnover as well as net growth.

Similarity is measured three ways: TF--IDF cosine over word 1--2-grams, word 5-gram Jaccard with
asymmetric containment, and MiniLM embeddings
(\code{sentence-transformers/all-MiniLM-L6-v2}, 256-token chunks, mean-pooled and L2-normalized).
Pairwise analyses use a 95-document representative subset (the longest file per
vendor$\times$archetype pair) to keep dendrograms legible while spanning all 62 vendors; all other
analyses use the full corpus. Because the block classifier supports the headline result, we run a
\emph{pilot validation}: 48 stratified blocks (six per predicted bucket, fixed seed) scored against a
gold functional label. This is automated scoring, not independent human coding, and six blocks per
bucket cannot support a confidence interval. Table~\ref{tab:val} reports precision. Legal/IP is 6/6;
injection boundary, identity/persona, and memory/context are 5/6; formatting/style, tool/protocol,
and safety policy are 4/6; other is 2/6. Overall precision is 35/48 (73\%). The tool/protocol share
therefore rests on a moderately noisy classifier; a larger human-coded sample remains necessary.

\begin{table}[t]
\centering
\caption{Block-classifier \emph{pilot} validation: a stratified sample of 48 blocks (six per
predicted bucket) from the full 407-file corpus. Per-bucket cells rest on $n{=}6$, the 73\% overall
on $n{=}48$; both directional.}
\label{tab:val}
\scriptsize
\begin{tabular}{lc}
\toprule
Predicted bucket & Precision \\
\midrule
Legal/IP           & 6/6 \\
Injection boundary & 5/6 \\
Identity/persona   & 5/6 \\
Memory/context     & 5/6 \\
Formatting/style   & 4/6 \\
Tool/protocol      & 4/6 \\
Safety policy      & 4/6 \\
Other (catch-all)  & 2/6 \\
\midrule
Overall            & 35/48 (73\%) \\
\bottomrule
\end{tabular}
\end{table}

\section{Results}
The corpus is heterogeneous. Chat-assistant prompts are the most numerous and longest group, while
coding-agent and autonomous-agent prompts contain dense operational instructions for tools, files,
shell access, and task loops (Table~\ref{tab:arch}, Fig.~\ref{fig:arch}).

\subsection{Textual and Structural Reuse and Semantic Association}
\label{sec:reuse}
\begin{figure}[t]
\centering
\includegraphics[width=0.65\columnwidth]{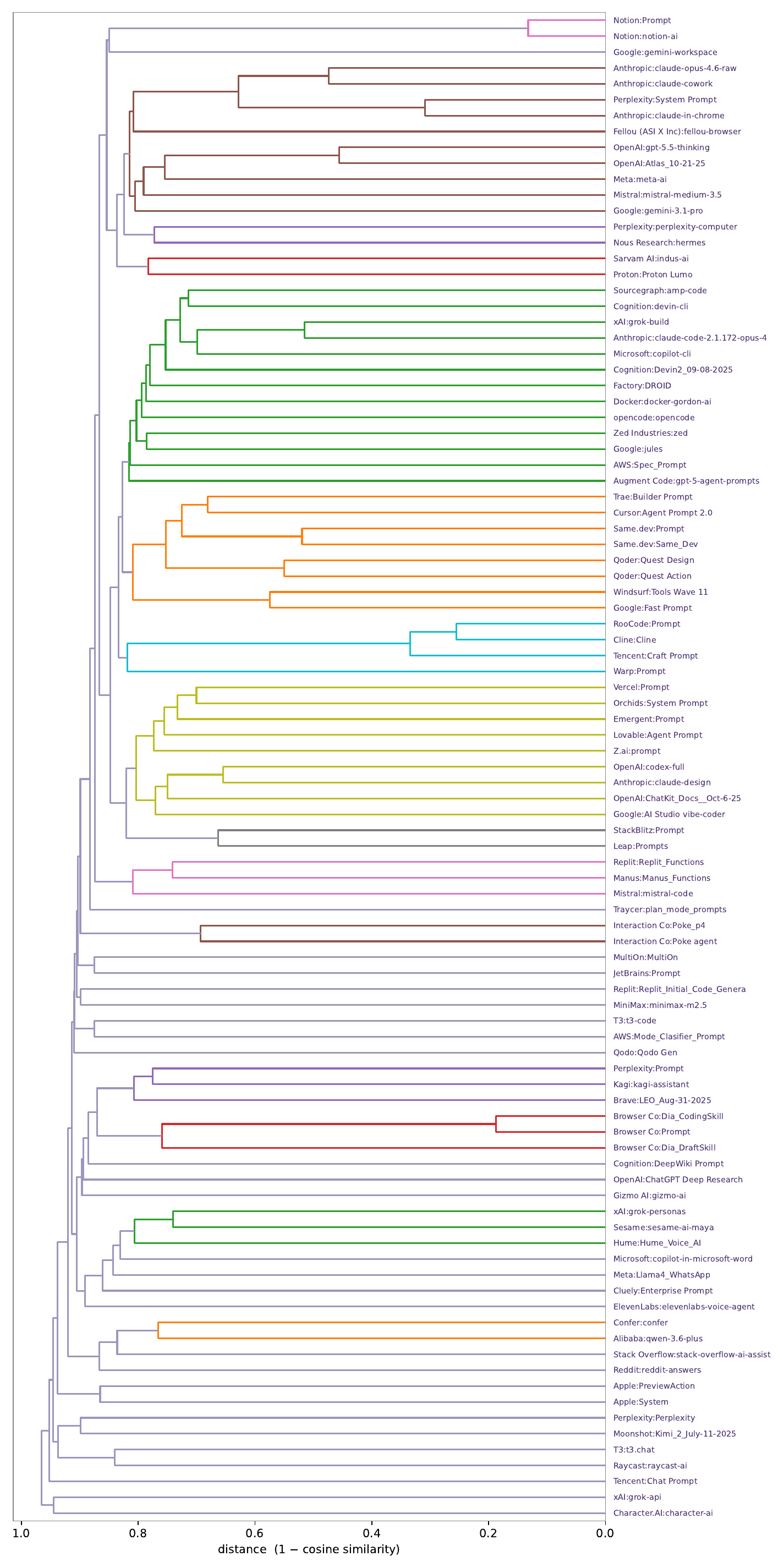}
\caption{Hierarchical clustering of 95 representative prompts (one per vendor$\times$archetype) by TF--IDF cosine similarity.}
\label{fig:tfidf}
\end{figure}

TF--IDF clustering over the 95-document representative subset shows tight same-vendor families and
cross-vendor coding-agent clusters (Fig.~\ref{fig:tfidf}). The highest cross-vendor pair is
Cline--RooCode (cosine 0.745), followed by Cline--Tencent CodeBuddy (0.725) and Anthropic's
Chrome-extension assistant--Perplexity's browser assistant (0.691). Cline--RooCode is a useful check:
RooCode is a public open-source fork of Cline, so it should rank highly. Tencent's CodeBuddy sits in
the same cluster (RooCode--Tencent cosine 0.605); its identity sentence is identical to Cline's and
RooCode's except for the assistant name, indicating shared source text. The Anthropic--Perplexity
pair is a close paraphrase rather than a verbatim copy, both organize browser-agent injection defense
around the same hierarchy, untrusted web/DOM/email content, and enumerated injection patterns. The
text cannot show whether this reflects directional influence or convergence on a common convention.

Literal reuse makes the copied-text relationships clearer. RooCode has 54\% of its word 5-grams
contained in Cline; Tencent CodeBuddy's Craft prompt has 50\%; and Anthropic's Chrome-extension
assistant and Perplexity's browser assistant share 40\% containment. Table~\ref{tab:reuse} lists the
highest cross-vendor literal-overlap pairs from the same subset.

\begin{table}[t]
\centering
\scriptsize
\caption{Highest cross-vendor pairs by literal overlap (95-document representative subset).}
\label{tab:reuse}
\begin{tabular}{lccc}
\toprule
Pair & TF--IDF & Jaccard & Max \\
 & cosine & &  containment \\
\midrule
Cline -- RooCode & 0.71 & 0.34 & 54\% \\
Cline -- Tencent & 0.69 & 0.29 & 50\% \\
Anthropic (Chrome) -- Perplexity & 0.68 & 0.20 & 40\% \\
RooCode -- Tencent & 0.53 & 0.17 & 31\% \\
Google -- Windsurf & 0.47 & 0.10 & 22\% \\
Cursor -- Same.dev & 0.33 & 0.08 & 20\% \\
Qoder -- Same.dev & 0.22 & 0.06 & 14\% \\
Trae -- Traycer & 0.23 & 0.06 & 20\% \\
Anthropic (Code) -- xAI (Grok) & 0.46 & 0.05 & 18\% \\
\bottomrule
\end{tabular}
\end{table}

\begin{figure}[t]
\centering
\includegraphics[width=0.7\columnwidth]{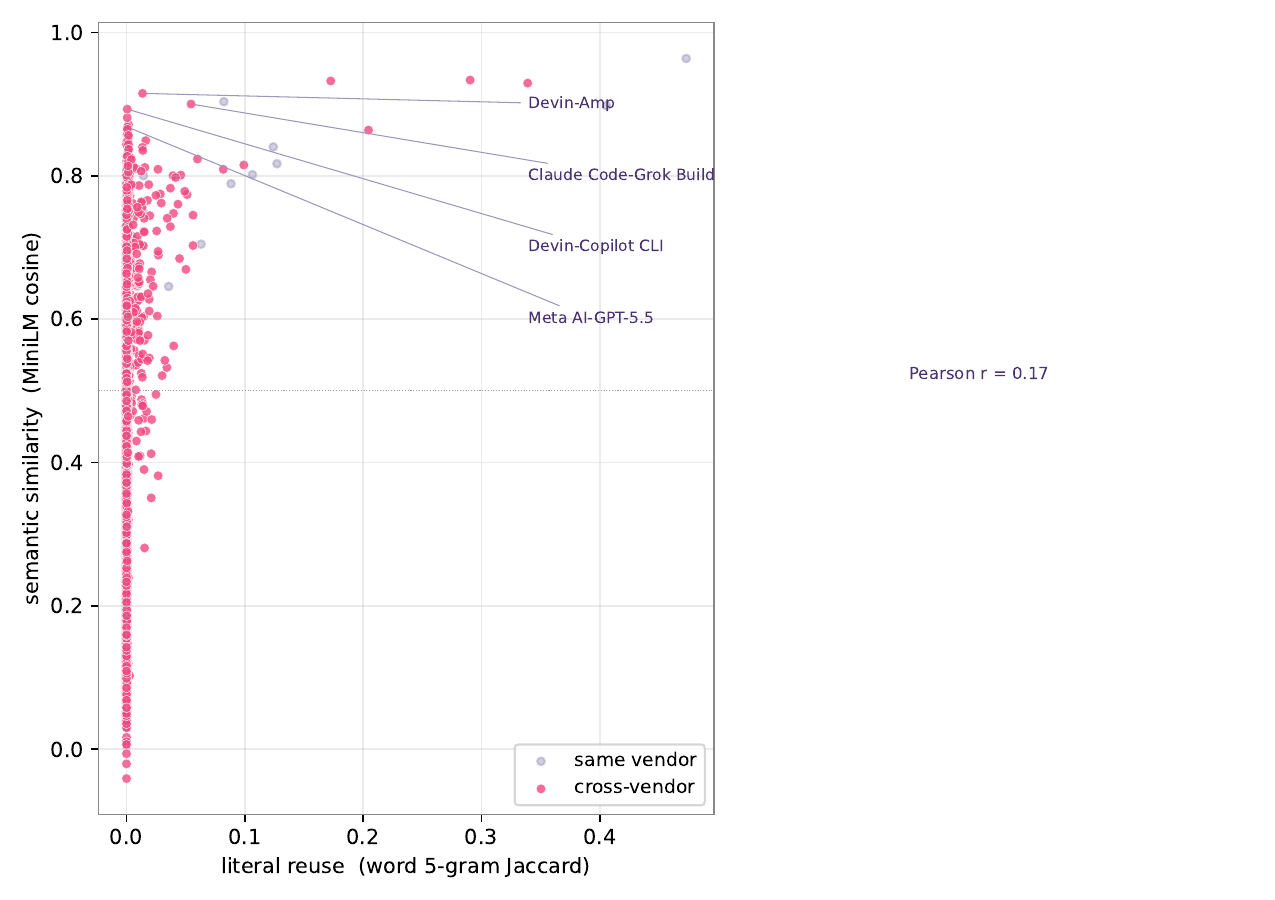}
\caption{Literal reuse versus semantic similarity for prompt pairs.}
\label{fig:litsem}
\end{figure}

The MiniLM cross-check separates literal reuse from shared task structure. Literal Jaccard and
semantic cosine correlate only weakly (Pearson $r=0.175$). Several cross-vendor pairs are
semantically close with near-zero literal overlap, especially autonomous coding-agent CLIs:
Devin--Amp (semantic 0.915, literal 0.014), Devin--GitHub Copilot CLI (0.893, 0.001), Augment
Code--Devin (0.881, 0.001), and Claude Code--Copilot CLI (0.871, 0.002). Meta AI and ChatGPT-5.5
also converge semantically (0.868) with essentially no shared text. Anthropic Chrome
extension--Perplexity is high on both axes (semantic 0.864, literal 0.20; Fig.~\ref{fig:litsem}).

% \subsection{ Reuse}
\label{sec:structural}
Literal 5-gram overlap and MiniLM cosine capture copied words and shared meaning, but not shared
architecture: XML tags, markdown headers, fenced code blocks, and list or key-value markers. We strip
natural language from each prompt and keep a structural skeleton, e.g., \code{<thinking>}, \code{H2},
\code{CODE}, or \code{UL}. Skeletons are compared by profile cosine (TF--IDF over skeleton
1--2-grams) and sequence-shingle Jaccard. To avoid trivial all-bullet matches, reported structural
matches are restricted to the 58 of 95 representative prompts with at least 12 tokens and 8 distinct
token types; structure-light prompts remain in the matrices but are not ranked as matches.

Two checks support this measure. A token-shuffle null leaves cross-vendor overlap almost unchanged
(mean Jaccard 0.038 ordered versus 0.031 shuffled), so structural convergence is about which elements
a prompt uses, not their sequence. We therefore report profile cosine. Structural-profile similarity
also correlates weakly with literal reuse ($r=0.14$) and semantic similarity ($r=0.04$), so it is not
just a proxy for the other two measures.

The strongest cross-vendor structural match is also the strongest pair overall. Anthropic's Chrome
extension and Perplexity's browser assistant rank first on all three measures (structural-profile
cosine 0.86, literal Jaccard 0.20, semantic cosine 0.86). Beyond that pair, the highest structural
matches mostly occur without literal copying: AWS Kiro spec-mode--Microsoft Word Copilot (0.73,
0.000), Augment Code--Google AI-Studio vibe-coder (0.68, 0.000), and Cluely--ElevenLabs voice agent
(0.67, 0.000). Literal copying and architectural convergence therefore describe different subsets of
files, with Anthropic--Perplexity as the exception where both hold. The full matrix does not support
one shared skeleton for all coding agents or all chat assistants; the structural matches mostly
suggest convergence on agent-design conventions rather than cross-vendor structural plagiarism.

\begin{figure}[t]
\centering
\includegraphics[width=0.7\columnwidth]{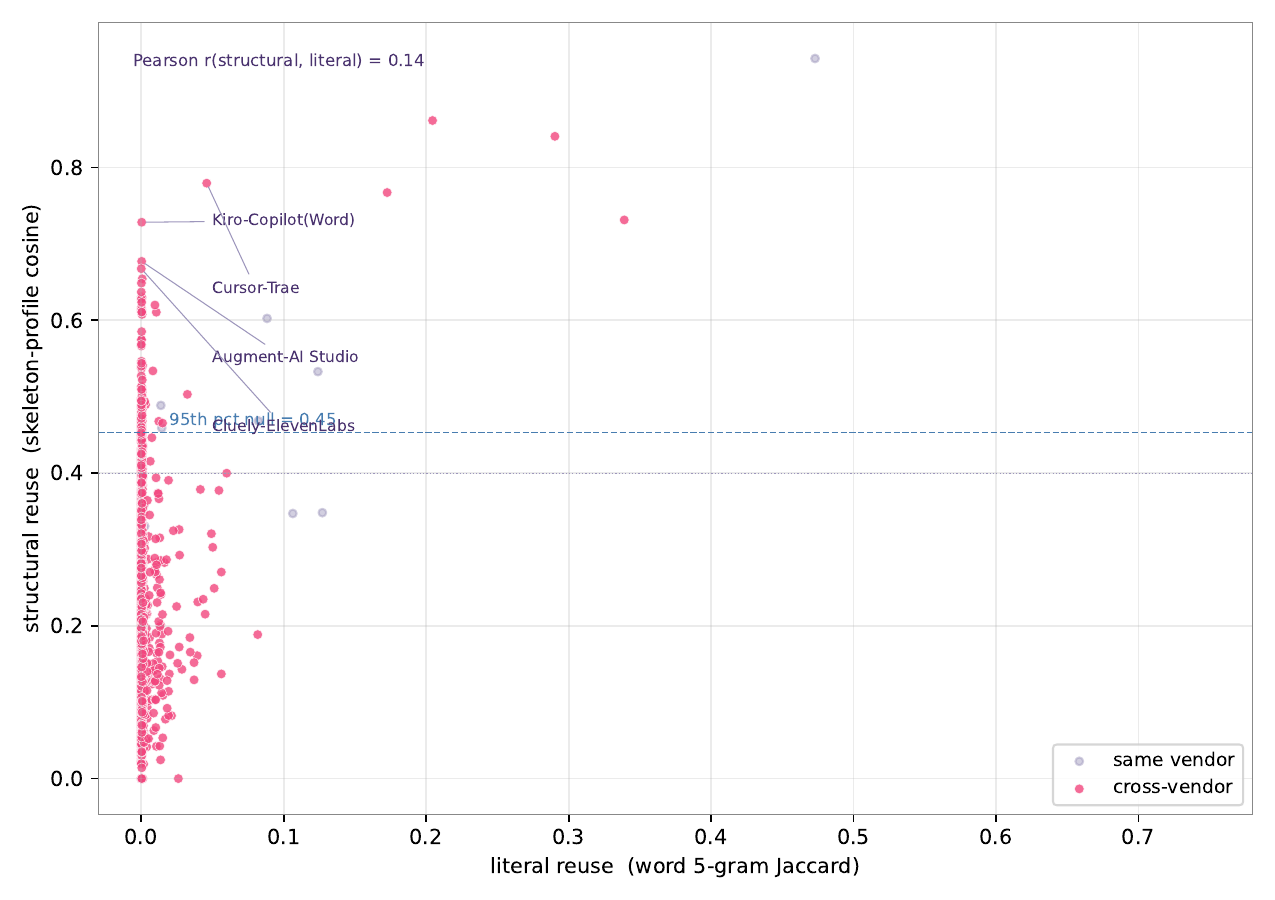}
\caption{Copied words (literal 5-gram Jaccard) versus shared architecture
(skeleton-profile cosine) for structure-bearing prompt pairs.}
\label{fig:struct}
\end{figure}

\subsection{Functional Composition and Capabilities}
\begin{figure}[t]
\centering
\includegraphics[width=0.7\columnwidth]{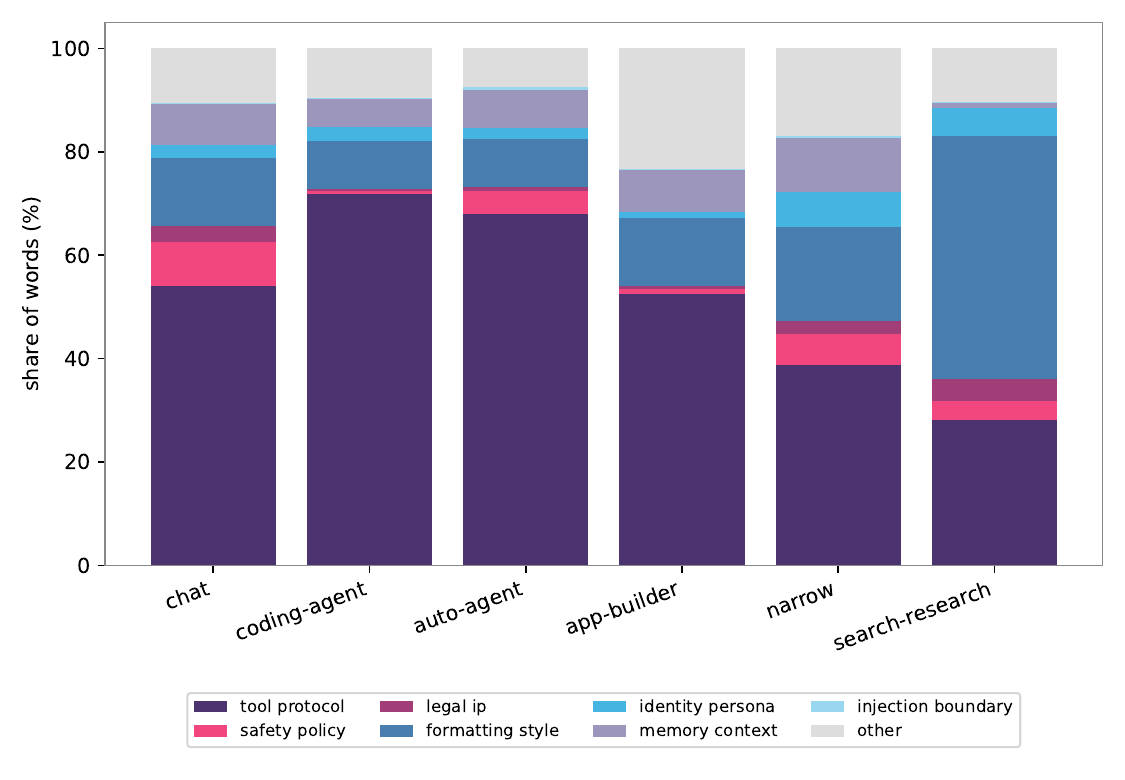}
\caption{Block-level token taxonomy by archetype.}
\label{fig:block}
\end{figure}

Block taxonomy provides a word-level denominator for the corpus. Under this noisy but auditable
classifier, 58.0\% of classified words are assigned to tool/protocol blocks, 12.4\% to
formatting/style, 7.4\% to memory/context, 5.4\% to safety policy, 2.9\% to identity/persona, 2.1\%
to legal/IP, 0.2\% to explicit injection-boundary language, and 11.7\% to other material. The
tool/protocol share spans chat-facing and narrow/fragment prompts as well as coding-agent tooling.
Coding agents and autonomous agents exceed 80\% tool/protocol content; search/research is the main
exception, with most words assigned to formatting and citation rules (Fig.~\ref{fig:block}).
Tool/protocol still exceeds every other bucket combined outside "other."

\begin{figure}[t]
\centering
\includegraphics[width=0.8\columnwidth]{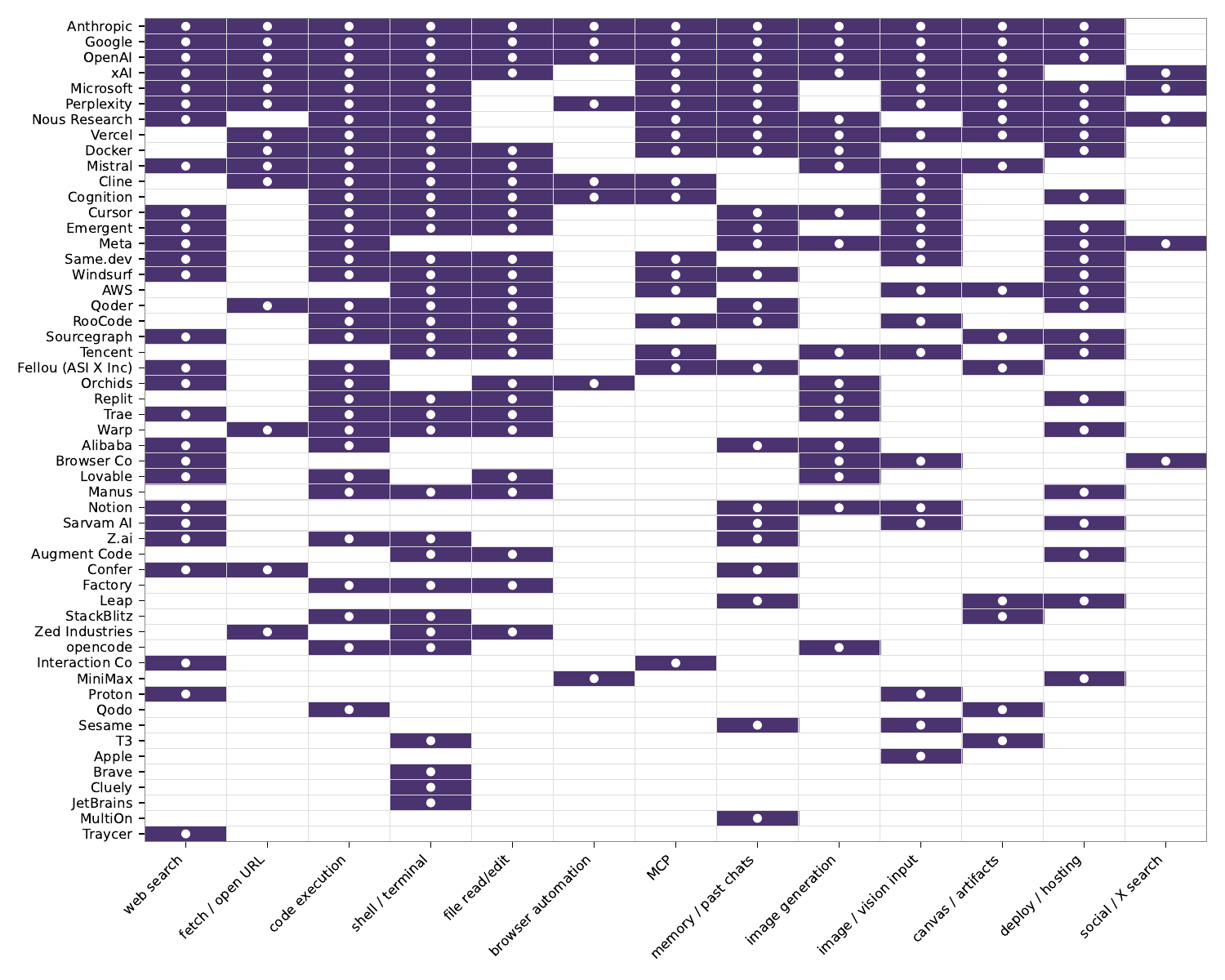}
\caption{Vendor by capability-family grid. A mark means the capability is described by at least one prompt from the vendor.}
\label{fig:caps}
\end{figure}

Thirteen capability families are detected lexically: web search, URL fetch/open, code execution,
shell/terminal, file read/edit, browser automation, MCP, memory/past chats, image generation, vision
input, canvas/artifacts, deploy/hosting, and social/X search. On the representative subset,
Anthropic, Google, and OpenAI each describe 12 of 13 families (all but social/X search, which is
xAI-specific), with xAI at 11 of 13. Fig.~\ref{fig:caps} marks described capabilities, not confirmed
deployment.

\subsection{Comparison with Officially Published Prompts}
\label{sec:official}
A vendor-published archive lets us check fidelity and locate where the operational content appears.
Anthropic's release-notes prompts (Section~\ref{sec:corpus}) yield five matched official/leaked
pairs: Opus 4.6, 4.7, 4.8, Sonnet 4.6, and Fable~5. We run the same block classifier and 5-gram
containment measure over each pair (Table~\ref{tab:official}).

Within this vendor family, all five pairs show the same pattern. Official prompts average 3{,}080
words (range 2{,}677--3{,}678) and are safety-led (mean 34.5\% safety policy, 17.1\%
tool/protocol). Matched leaked counterparts average 20{,}215 words (range 14{,}190--24{,}784, mean
6.6$\times$ expansion) and invert the balance (57.2\% tool/protocol, 8.9\% safety). Containment of
the official text inside the leaked file is high for four pairs (76--94\%); Sonnet 4.6 is an outlier
at 38\%, so we flag rather than aggregate it. In these pairs, leaked files contain most of the
published behavioral core plus roughly 17{,}000 additional words of tool and protocol material absent
from public releases. Public prompts therefore look more safety-focused than their observed leaked
counterparts.

\begin{table}[t]
\centering
\scriptsize
\caption{Block taxonomy and containment of five officially published vs.\ leaked Claude prompt
pairs (\% of classified words; containment is the share of official 5-grams found verbatim in the
leaked file).}
\label{tab:official}
\begin{tabular}{lrrrr}
\toprule
Model & Off. & Leaked & Ratio & Contain. \\
\midrule
Fable 5     & 2{,}677 & 17{,}501 & 6.5$\times$ & 88\% \\
Opus 4.6    & 2{,}819 & 14{,}190 & 5.0$\times$ & 86\% \\
Opus 4.7    & 3{,}678 & 21{,}177 & 5.8$\times$ & 76\% \\
Opus 4.8    & 3{,}344 & 24{,}784 & 7.4$\times$ & 94\% \\
Sonnet 4.6  & 2{,}882 & 23{,}425 & 8.1$\times$ & 38\% \\
\midrule
Mean & 3{,}080 & 20{,}215 & 6.6$\times$ & 76\% \\
\bottomrule
\end{tabular}

\medskip
\begin{tabular}{l}
Tool/protocol \%: official mean 17.1, leaked mean 57.2 \\
Safety policy \%: official mean 34.5, leaked mean 8.9 \\
\end{tabular}
\end{table}

\subsection{Safety, Legal, and Injection Risk}
\begin{figure}[t]
\centering
\includegraphics[width=\columnwidth]{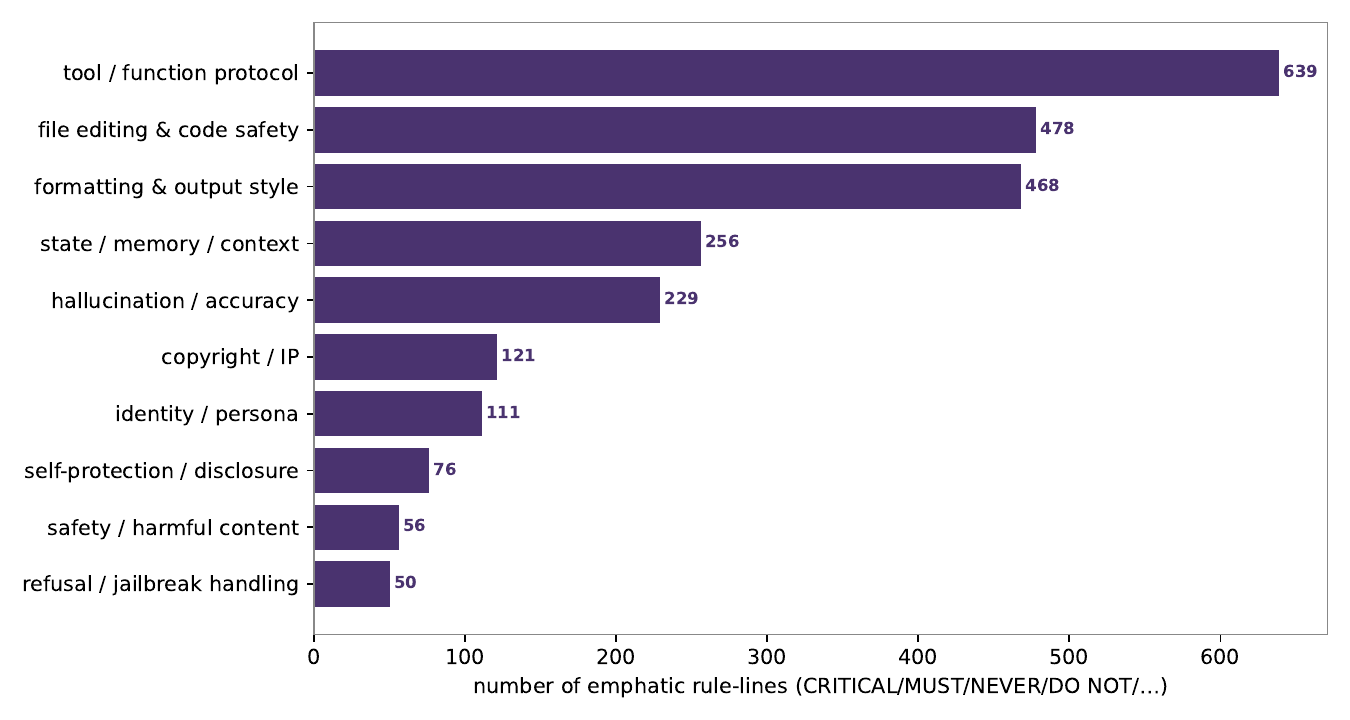}
\caption{Emphatic rule-lines by category. A line may match multiple categories.}
\label{fig:failure}
\end{figure}

The strictest language guards operational behavior, not harmful content. Among the 4{,}725 rule-lines
with emphatic markers such as \code{CRITICAL}, \code{IMPORTANT},
\code{MUST NOT}, \code{NEVER}, \code{DO NOT}, \code{STRICTLY}, or \code{REQUIRED}, the most
frequent categories are tool/function protocol (639), file-editing and code safety (478),
formatting/output style (468), state/memory/context (256), hallucination/accuracy (229),
copyright/IP (121), identity/persona (111), self-protection/non-disclosure (76), and
safety/harmful content (56), so the most forceful instructions cluster on tool use, file safety, and
formatting, with harmful-content rules more than an order of magnitude rarer than tool-protocol
rules (11.4:1; Fig.~\ref{fig:failure}).

\begin{figure}[t]
\centering
\includegraphics[width=\columnwidth]{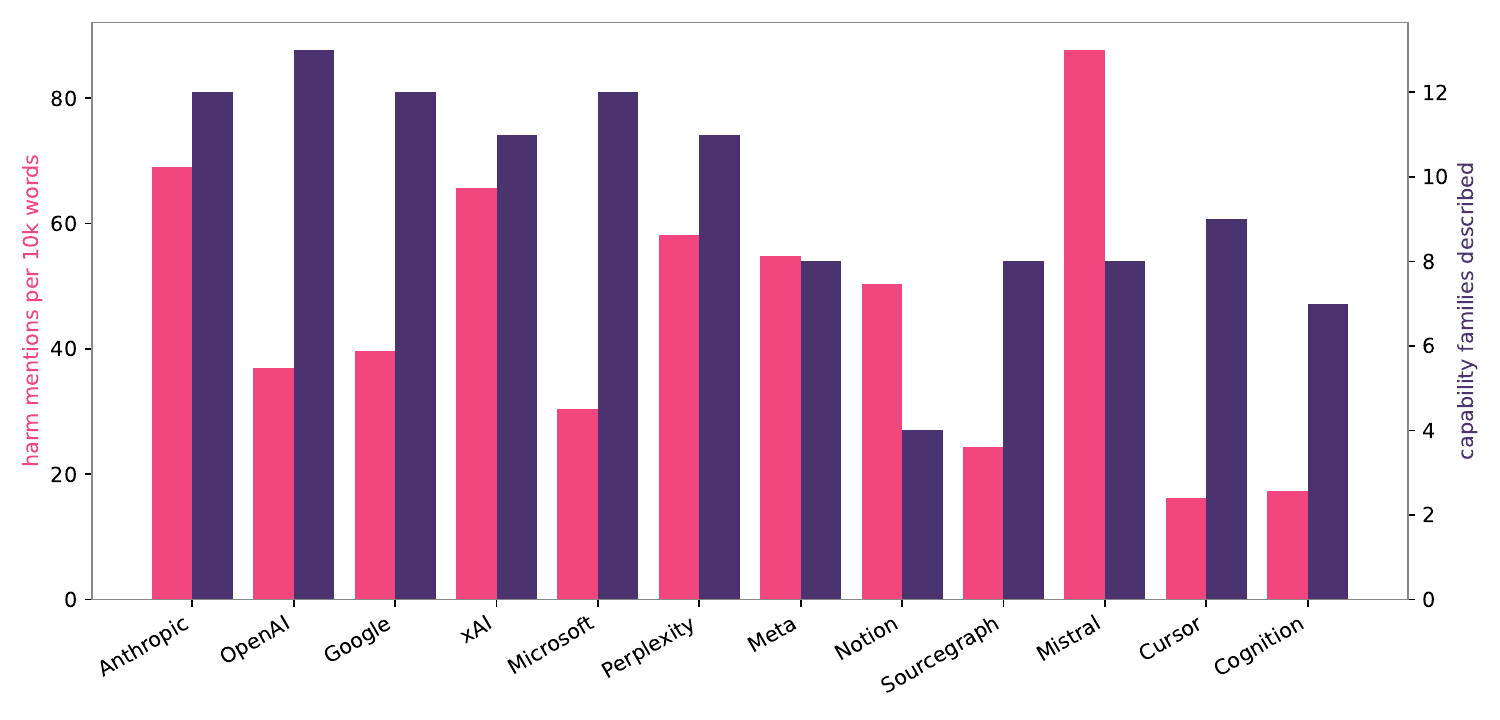}
\caption{Normalized harm-mention rate and capability breadth for the largest raw harm-count vendors.}
\label{fig:norm}
\end{figure}

Raw harm-category counts differ by vendor: Anthropic has 6{,}192 mentions across 97 coded files,
OpenAI 1{,}042 across 113, Google 264, and xAI 241. Word-normalized rates compress the gap. Among
vendors with at least three files, Mistral has the highest rate (87.7 mentions per 10{,}000 words, on
3 files), followed by Anthropic (69.0), xAI (65.7), Perplexity (58.2), Meta (54.9), Proton (49.9),
and OpenAI (37.0). Anthropic's distinctiveness is total prompt length and category breadth rather
than uniquely high per-token density; rates on 3-file vendors are noisy (Fig.~\ref{fig:norm}).

\begin{figure}[t]
\centering
\includegraphics[width=0.9\columnwidth]{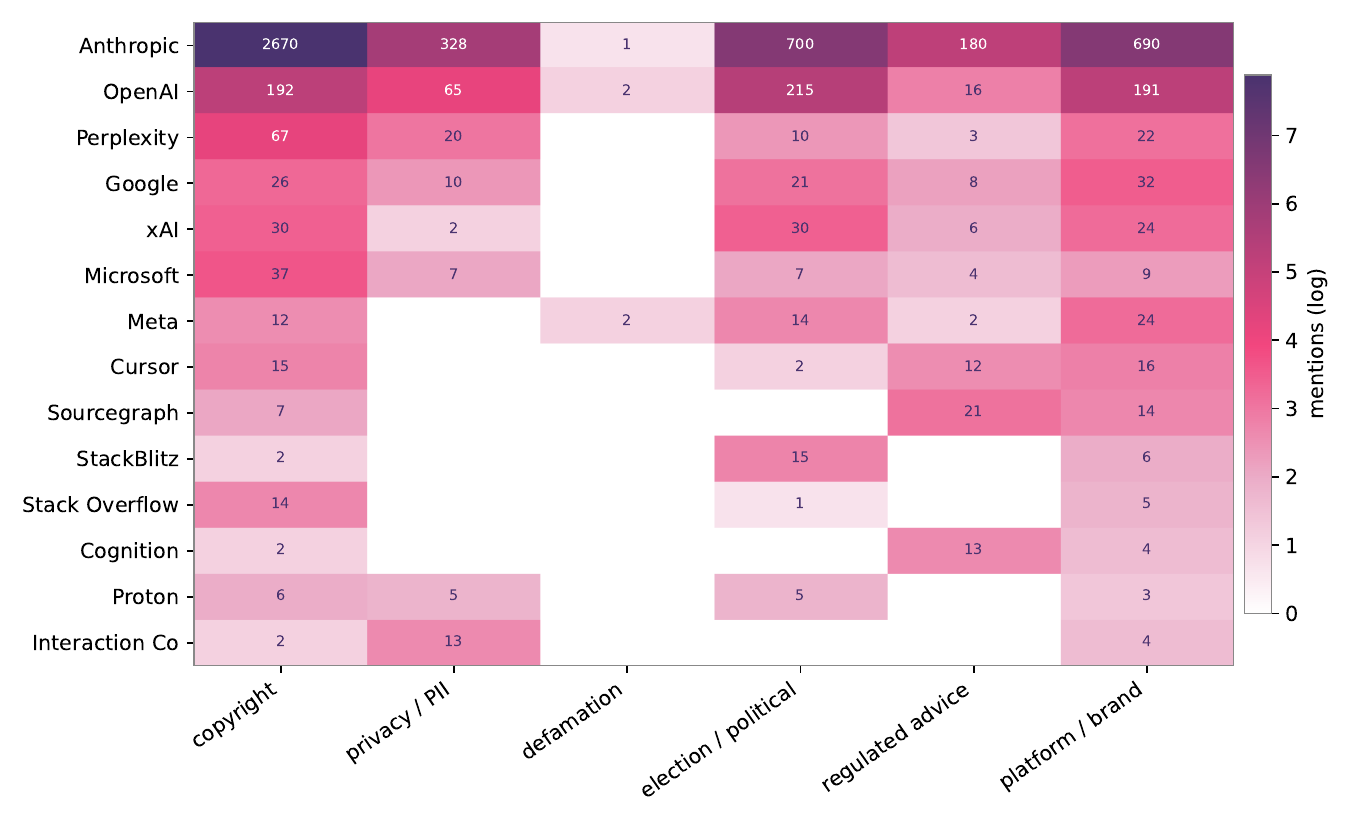}
\caption{Legal-risk terms by vendor. Counts are lexical and include copyright, privacy/PII, defamation,
election/political, regulated-advice, and platform/brand terms.}
\label{fig:legal}
\end{figure}

Legal-risk terms are also concentrated. The lexicon counts 4{,}569 mentions for Anthropic, 681 for
OpenAI, 122 for Perplexity, 97 for Google, 92 for xAI, and lower totals for other vendors. Copyright
dominates (3{,}157 mentions), followed by platform/brand (1{,}103) and election/political language
(1{,}048), ahead of privacy/PII (467) and regulated advice (296). This ordering is driven mainly by
OpenAI and Anthropic chat-persona and safety-policy fragments (Fig.~\ref{fig:legal}).

Non-disclosure instructions appear often, while stronger trust-boundary patterns are less common. The
detector distinguishes non-disclosure, precedence hierarchies, data-only tagging,
ignore-embedded-instructions rules, and explicit jailbreak refusal. In the representative subset, 35
of 62 vendors (56\%) show at least one technique; most rely on non-disclosure alone, and only seven
(Table~\ref{tab:inj}) combine three or more. Six vendors (Brave, Docker, Microsoft, Nous Research,
Perplexity, Sesame) combine exactly two. Notion relies only on a precedence hierarchy, and Vercel
only on data-only tagging. Because the corpus itself mostly consists of leaks, non-disclosure should
not be read as an effective boundary mechanism.

\begin{table}[t]
\centering
\scriptsize
\caption{Prompt-injection defense techniques for the 7 vendors (of 35 with $\geq$1 technique) that
combine three or more. ND: non-disclosure of prompt; PH: precedence hierarchy; DT: data-only tagging;
IE: ignore embedded instructions; JR: explicit jailbreak refusal.}
\label{tab:inj}
\begin{tabular}{lccccc}
\toprule
Vendor & ND & PH & DT & IE & JR \\
\midrule
Anthropic     & \checkmark & \checkmark & \checkmark & \checkmark &            \\
Cognition     & \checkmark & \checkmark & \checkmark & \checkmark &            \\
xAI           & \checkmark & \checkmark &            & \checkmark & \checkmark \\
Browser Co    & \checkmark &            & \checkmark & \checkmark &            \\
Google        & \checkmark & \checkmark &            & \checkmark &            \\
OpenAI        & \checkmark & \checkmark &            & \checkmark &            \\
Sourcegraph   & \checkmark & \checkmark &            & \checkmark &            \\
\bottomrule
\end{tabular}
\end{table}

\subsection{Version Growth, Instruction Tensions, and Autonomy}
\label{sec:version}
\begin{figure*}[t]
\centering
\begin{subfigure}[t]{0.31\textwidth}\centering
\includegraphics[width=\linewidth]{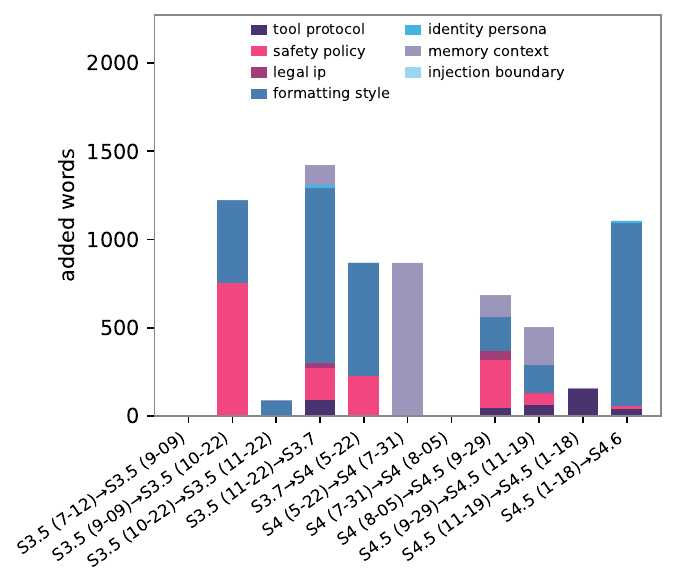}\caption{Sonnet (Off.)}\end{subfigure}\hfill
\begin{subfigure}[t]{0.31\textwidth}\centering
\includegraphics[width=\linewidth]{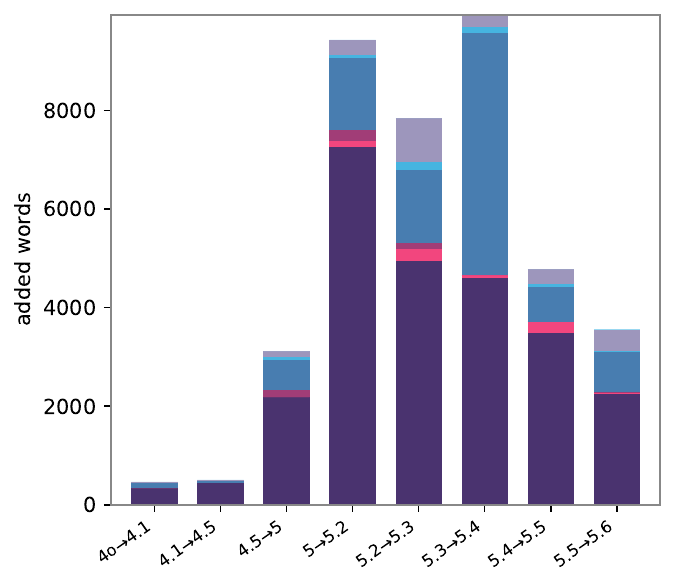}\caption{ChatGPT}\end{subfigure}\hfill
\begin{subfigure}[t]{0.31\textwidth}\centering
\includegraphics[width=\linewidth]{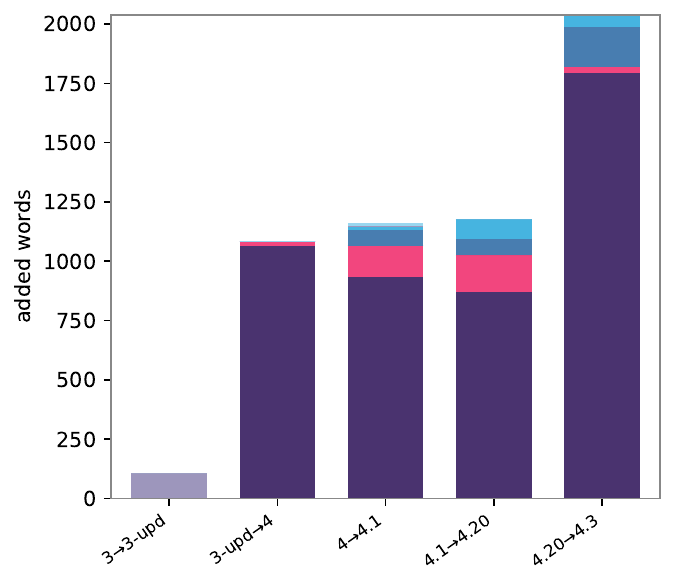}\caption{Grok}\end{subfigure}
\caption{Consecutive version diffs, added words by block type; legend in panel (a) applies to all
three (three of the five chains; Opus-Official
and legacy-leaked Claude chain numbers are in the text).}
\label{fig:verdiff}
\end{figure*}

A verified-date official chain lets us separate published prompt growth from apparent drift in
leak-derived snapshots. The two disagree sharply. The official Claude Sonnet chain has 12 dated
releases (11 steps) with mean per-step turnover of 1{,}319 words added$+$removed (mean net change
362). The official Opus chain has 9 releases (8 steps) and is similar (turnover 1{,}424, net 585).
A leaked-only Claude chain (CL4R1T4S) over a similar span shows more than $10\times$ that turnover:
mean 15{,}049 words added$+$removed per step, mean net 2{,}878. Its largest step, Opus 4.6 to Opus
4.7, adds 16{,}946 words and removes 9{,}959; 10{,}262 added words are tool/protocol and 1{,}574
legal/IP. ChatGPT's leaked chain is also high (mean 9{,}632 words/step), while Grok's is lower
(1{,}940). We treat official chains as documented prompt evolution, and leaked chains as evidence
about instability in the leak record, not direct production change (Fig.~\ref{fig:verdiff}).

Positive imperatives outnumber negative imperatives across all archetypes. Coding agents have the
lowest negative-imperative share (29.0\%), followed by app-builders and autonomous agents (29.1\%,
29.9\%); narrow/embedded assistants are highest (36.9\%), and chat assistants sit at 31.7\%. A
tension detector flags co-occurring instruction pairs: 286 files contain both conciseness and
long-form/thoroughness language, 116 combine list bans with markdown/table/list requirements, 85
contain both ask-clarifying-question and do-not-ask language, and 15 contain both safety-policy and
unrestricted-content language. These flags do not prove local contradiction.

\begin{figure}[t]
\centering
\includegraphics[width=0.8\columnwidth]{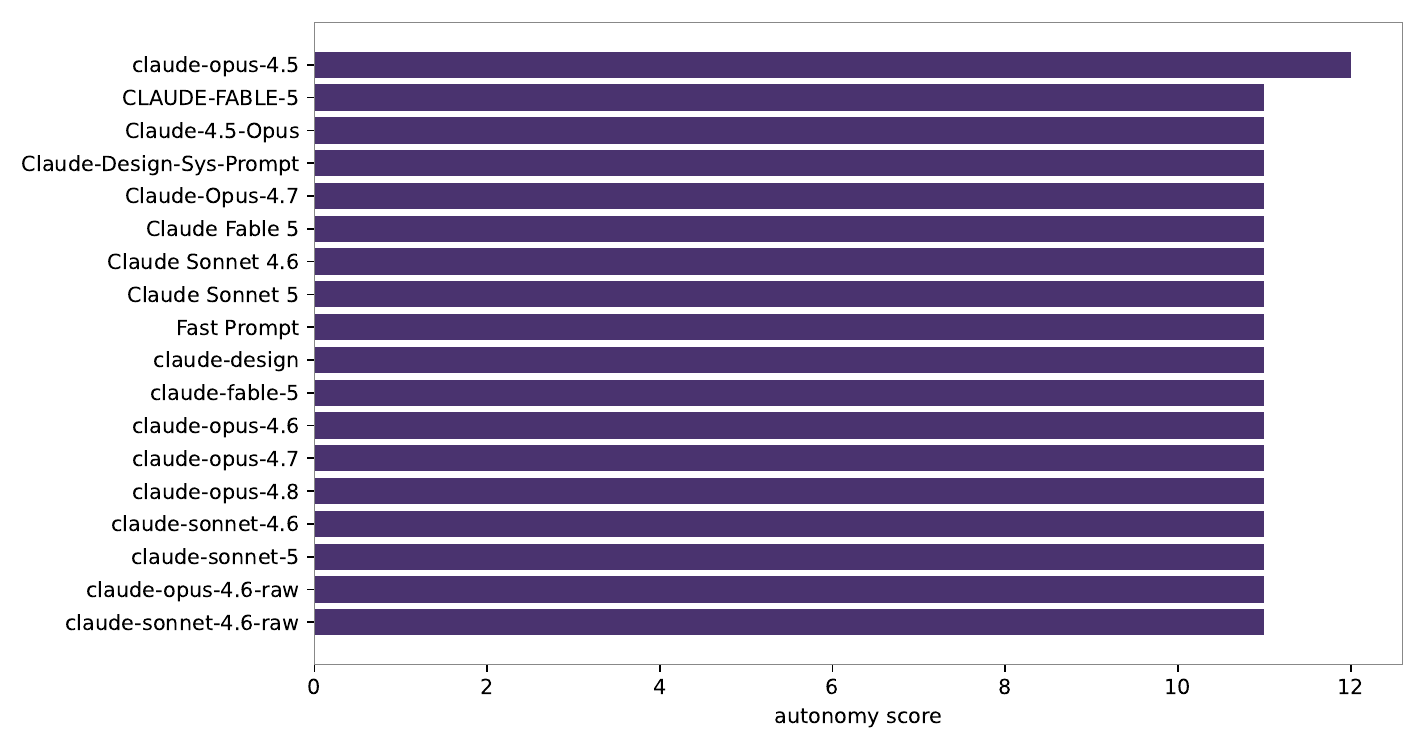}
\caption{Highest tool-autonomy scores. Approval-boundary language subtracts one point.}
\label{fig:auto}
\end{figure}

The tool-autonomy index scores prompts for shell access, file editing, browser control, web access,
code execution, deployment, memory, planning loops, and approval boundaries. Anthropic chat prompts
still dominate the top of the ranking (18 of the top 25 scores), joined by Cognition (Devin), Manus,
Replit, Google's Antigravity agent, and several newly added agentic CLIs (Emergent, Nous Research).
This index measures described action permissions, not confirmed runtime affordances (Fig.~\ref{fig:auto}).

\begin{figure}[t]
\centering
\includegraphics[width=\columnwidth]{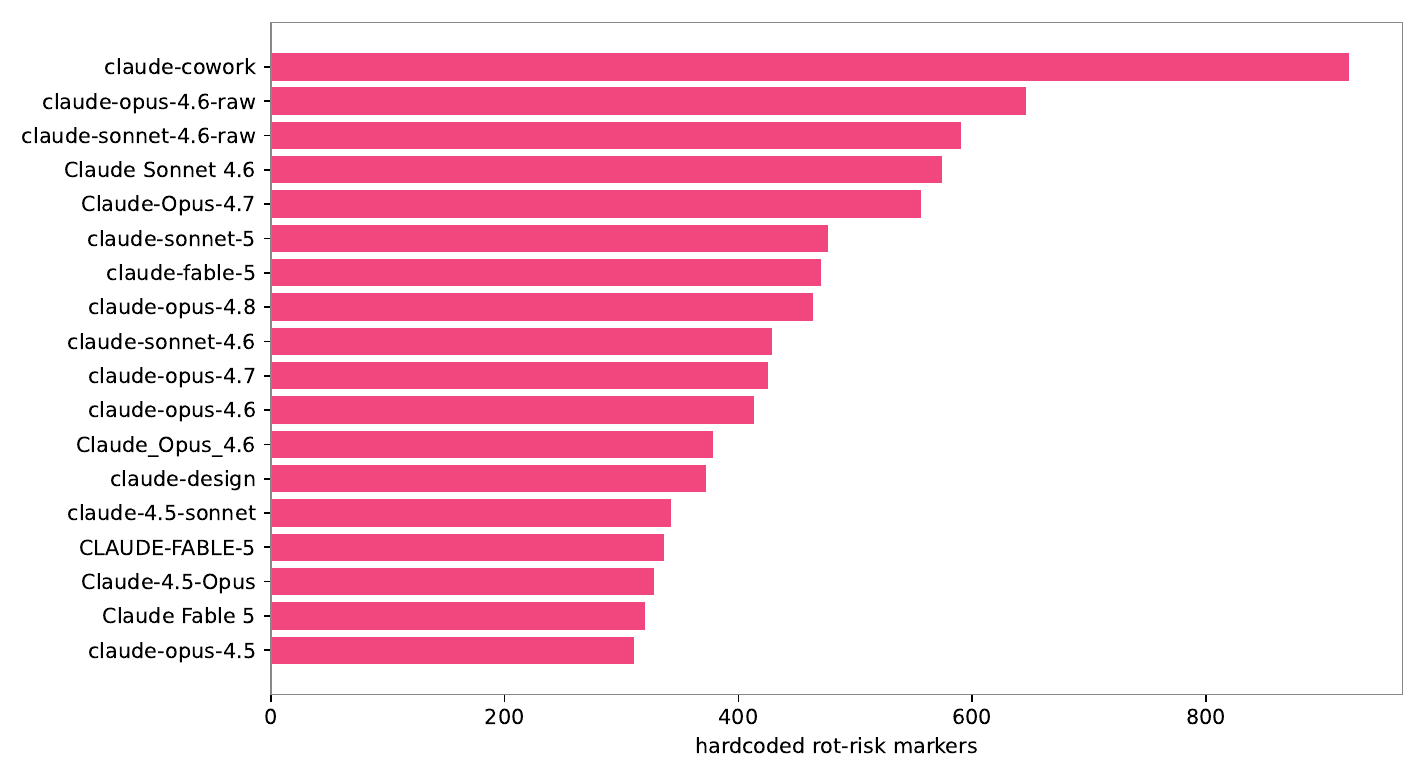}
\caption{Hardcoded rot-risk markers by prompt.}
\label{fig:rot}
\end{figure}

Hardcoded rot-risk markers concentrate in Anthropic chat and auto-agent prompts. Anthropic's Cowork
auto-agent prompt (\code{claude-cowork.md}, 34{,}900 words) carries 922 markers, ahead of
\code{claude-opus-4.6-raw} (646), \code{claude-sonnet-4.6-raw} (590), Claude Sonnet 4.6 (574), and
Claude Opus 4.7 (556) (Fig.~\ref{fig:rot}). The detector counts URLs, dates, version strings, model
identifiers, and CDN/package references. Raw totals include repetition: in \code{claude-cowork.md},
753 model-identifier hits collapse to 15 distinct identifiers and 12 URL hits to 6 distinct URLs.
Even deduplicated, a prompt with over a dozen perishable references has many places to age silently.

\subsection{Refusal Semantics}
\begin{figure}[t]
\centering
\includegraphics[width=\columnwidth]{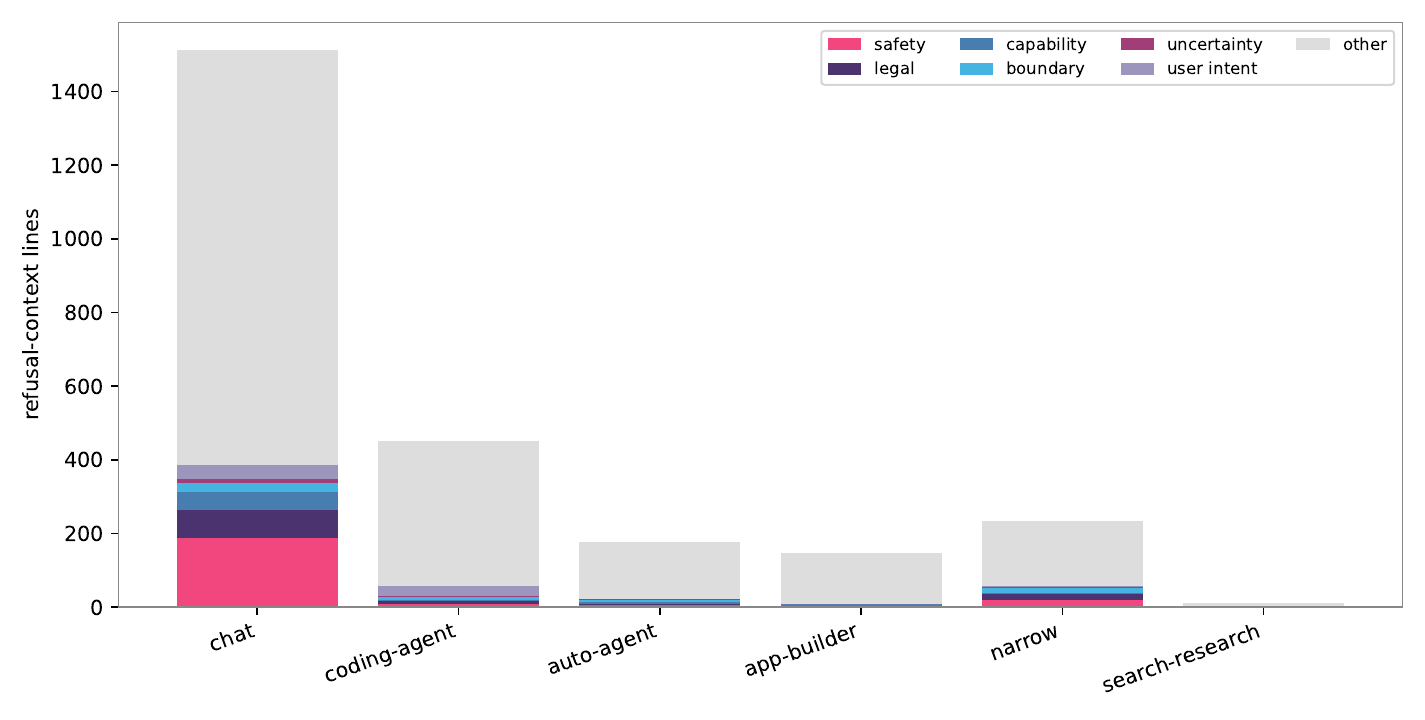}
\caption{Refusal semantics by archetype.}
\label{fig:refusal}
\end{figure}

Refusal-context lines are identified by refusal or inability language and bucketed by nearby terms.
Most remain uncategorized by the current lexicon (1{,}998), followed by safety (220), legal (110),
user-intent/abuse (67), capability limits (63), boundary protection (54), and uncertainty (20).
Refusal instructions exist, but are not the largest measured category under the current detectors
(Fig.~\ref{fig:refusal}).

\subsection{External-Tool Cross-Check}
\label{sec:crosscheck}
As a check on our detectors, we ran two off-the-shelf analyzers over all 407 prompts: PromptLint, a
rule-based static analyzer for LLM prompts~\cite{promptlint}, and textstat, a classical readability
library~\cite{textstat}. Neither shares our regular expressions, and textstat shares none of
PromptLint's rules.

Both tools agree with the operational and maintenance-debt findings. PromptLint's quality dimension
is its weakest score (mean 57.8--74.4 of 100 across archetypes, against 88.1--100 for completeness),
and its most common rules match the paper's account: over-length sentences fire on 95\% of prompts
(386/407), vague quantifiers on 87\%, inconsistent terminology on 81\%, and weak/passive phrasing on
78\%. textstat also rates the corpus as dense: Gunning-fog 12.8--15.4 with Flesch reading-ease in
the ``difficult'' 39--49 band.

\begin{figure}[t]
\centering
\includegraphics[width=\columnwidth]{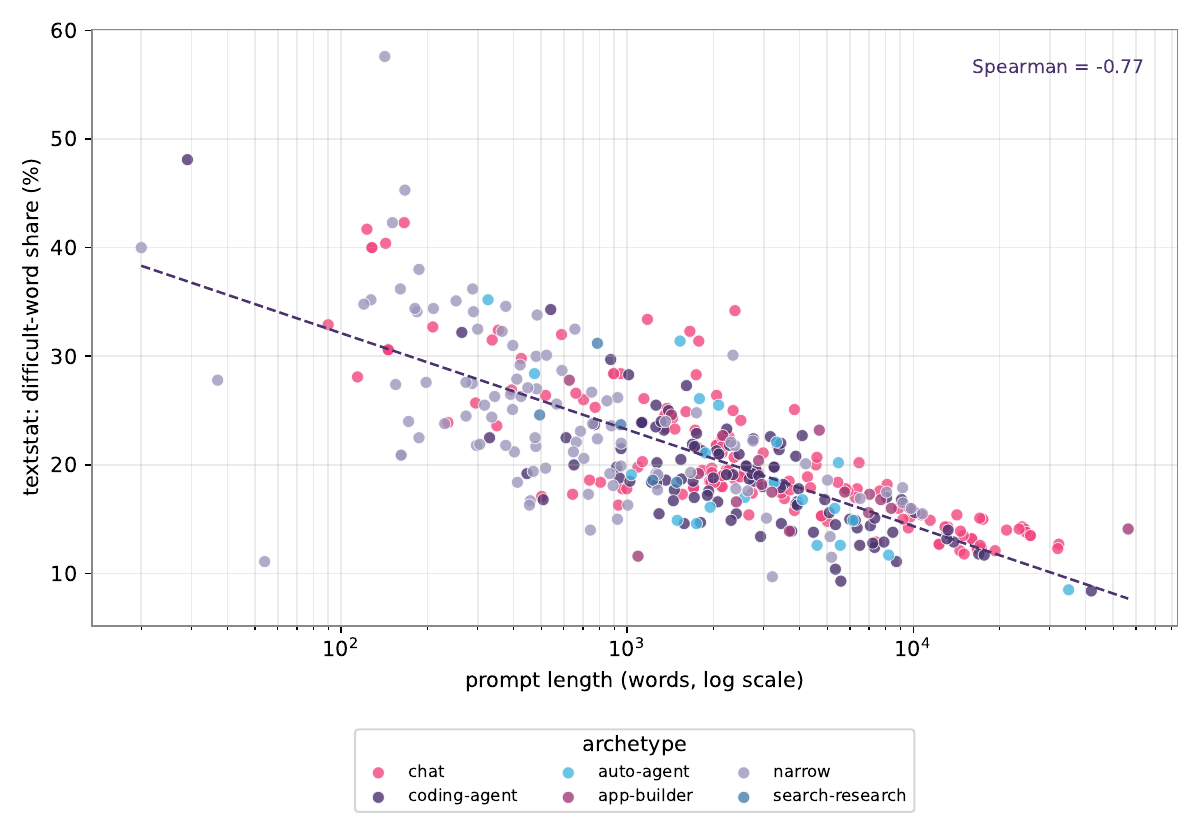}
\caption{Difficult-word share versus prompt length (textstat).}
\label{fig:lexlen}
\end{figure}

The cross-check also adds two caveats. First, PromptLint's security rules are intent-blind. Its
\code{secret-in-prompt} rule fired on 4 of 407 files (an Emergent prompt, an Orchids.app prompt, and
two OpenAI Codex files). Still, manual inspection found no credentials, only instructions about bearer
tokens and environment variables. App-builder prompts also score lowest on PromptLint security (mean
74.2), with chat assistants close behind (79.4), because both contain defensive, placeholder-laden,
and quoted-attack language. Second, PromptLint's long-sentence count is nearly collinear with
document length (Pearson 0.99), and agrees only weakly with textstat's sentence model (Spearman
0.20), because lists, code, and XML are segmented differently. Generic text tools transfer
imperfectly to prompts because the documents are not prose.

textstat adds one finding: difficult-word share falls as prompts grow (Spearman $-0.77$), from
roughly 30\% in short prompts to 15\% in the longest (Fig.~\ref{fig:lexlen}). The added length is
repetitive, plain-vocabulary tool text rather than denser prose, supporting the operational reading
through a method unrelated to our block classifier.

\section{Discussion and Limitations}
The analyses point to three conclusions: leaked prompts are mostly operational specifications, reuse
and convergence occur through different mechanisms, and maintenance costs are visible in both
verified prompt histories and leak-derived snapshots.

First, these prompts are primarily operational documents. Tool-use protocols, file-editing rules,
output formats, browsing constraints, and execution instructions account for most words and emphatic
rules; safety and legal language are smaller. This does not make safety instructions unimportant--a
single sentence can carry decisive policy weight--but it means leaked prompts should not be read
mainly as values documents. For security analysis, prompt extraction exposes mostly product and
tooling logic~\cite{DBLP:journals/corr/abs-2211-09527,zhang2024effective}, and indirect injection
mostly targets operational directives rather than refusal rules~\cite{DBLP:conf/ccs/AbdelnabiGMEHF23,DBLP:conf/nips/DebenedettiZBB024,DBLP:journals/corr/abs-2506-08837}.
The Anthropic matched pairs show the same pattern: the public prompt is safety-led, while the leaked
counterpart wraps that core in much more tool infrastructure (mean 6.6$\times$). Extending this
result requires comparable archives from other vendors.

Second, reuse and convergence occur at different levels. Same-vendor versions cluster tightly.
Cross-vendor semantic similarity largely reflects shared product roles: Devin, Amp, Copilot CLI, and
Augment Code form a tight semantic cluster with essentially zero literal overlap. Literal containment
is the evidence for verbatim transfer, concentrated in the Cursor/Windsurf/Same.dev cluster and the
Cline/RooCode fork, with Tencent's CodeBuddy in the same lineage. Structural comparison adds a third
case: the strongest cross-vendor structural convergence mostly has near-zero literal overlap,
consistent with shared design conventions rather than structural plagiarism. Anthropic's Chrome
extension and Perplexity's browser assistant are the exception, converging on words, meaning, and
architecture at once.

Third, prompt maintenance looks like a software-engineering problem. Official Anthropic release notes
show modest per-release turnover (mean 1{,}300--1{,}400 words added$+$removed), while the leaked-only
chain for the same product family shows more than $10\times$ that turnover, concentrated in
tool/protocol and legal/IP blocks. The gap warns against treating leak-derived version chains as
production histories. Rot-risk markers show prompts embedding perishable references (dates, model
identifiers, dependency versions, URLs), and tension flags show recurring conciseness/thoroughness and
ask/do-not-ask co-occurrences. Versioning, regression tests, and linting for stale references and
contradictory directives are therefore natural controls. The external-tool cross-check supports this
for quality, but not safety. For instance, PromptLint catches verbosity and inconsistency, yet its security rules
misfire on defensive and operational text.

\label{sec:threats}
The primary limitation is provenance. Most files are adversarial and unverified; they may be partial,
reconstructed, or modified, and we do not claim to measure official vendor policy. The matched-pair
check tempers this for five Anthropic documents (mean 76\% containment; Sonnet 4.6 is 38\%), but no
comparable vendor-published archive appears in the four source repositories. The multi-repository
merge adds another uncertainty as 341 of 407 files have no cross-repository duplicate. Detector
validity is also limited. Table~\ref{tab:val} puts tool/protocol and safety policy at 4/6, and the
audit was keyword-scored rather than independently human-coded. Manual archetype and posture labels
are single-coder judgments. Capability and autonomy detections show that a prompt describes a tool or
permission, not that it was active in production. Finally, the analysis is textual; whether these
prompt differences change deployed behavior remains open.

\section{Conclusion}
\label{sec:conclusion}
This paper analyzes a system-prompt corpus merged from four community collections. The observed
texts are dominated by operational specification: tool protocols, formatting constraints,
file/code-safety rules, capability descriptions, and hardcoded references. Safety, legal, refusal,
and injection-boundary instructions are present, but occupy a smaller measured share than operational
text. A leaked prompt is therefore better read as configuration-like text than as a statement of
values.

Five matched Anthropic releases show a 5--8$\times$ expansion and safety-to-tooling inversion between
public prompts and observed leaked counterparts. Because no other vendor here publishes a comparable
dated archive, that result calibrates rather than generalizes the corpus-wide findings. The corpus
also shows tight same-vendor version families, literal reuse concentrated among coding-agent and
browser-assistant prompts, and leak-derived version drift that should not be trusted without a
verified-date baseline. For maintenance, prompts should be versioned, regression-tested, and linted
for stale references and contradictory directives. For security, red-teaming should target tool-call
and data-handling boundaries, not only conversational refusal behavior. Future work should validate
manual labels with inter-rater agreement, separate permissive from prohibitive safety mentions, add
more official/leaked matched pairs, and test whether prompt-level differences predict model behavior.

\section*{Acknowledgment}
This work was partially supported by the European Commission under the Horizon Europe Programme, as part of the projects SAFEHORIZON (Grant Agreement No. 101168562) and CYMEDSEC (Grant Agreement No. 101094218). 

The content of this article does not reflect the official opinion of the European Union. Responsibility for the information and views expressed therein lies entirely with the authors.

This research has been partially funded by the University of Piraeus Research Center (UPRC).

\bibliographystyle{plain}
\bibliography{refs}

\end{document}